\documentclass{webofc}
\usepackage[varg]{txfonts}   
\begin{document}
\title{Heaven and Earth: Nuclear Astrophysics after GW170817}

\author{\firstname{Jorge} \lastname{Piekarewicz}\inst{1}\fnsep\thanks{\email{jpiekarewicz@fsu.edu}}}

\institute{Department of Physics, 
              Florida State University, 
              Tallahassee, FL 32306-4350, 
              USA}

\abstract{The historical detection of gravitational waves from the binary neutron star merger 
              GW170817 is providing fundamental new insights into the astrophysical site for the 
              creation of the heaviest elements in the cosmos and on the equation of state of 
              neutron-rich matter. Shortly after this historical detection, electromagnetic observations 
              of neutron stars together with measurements of the properties of neutron-rich nuclei at 
              terrestrial facilities have placed additional constraints on the dynamics of neutron-rich 
              matter. It is this unique synergy between heaven and earth that is the focus of this 
              article.} 

\maketitle

\section{Introduction}
\label{sec:Introduction}
More than a century ago, on November 25, 1915, Albert Einstein published his landmark paper 
on \emph{``The Field Equations of Gravitation''}\,\cite{Einstein:1915}. Only three years later he 
would unveiled one of the most remarkable predictions of the theory: the existence of gravitational
waves\,\cite{Einstein:1918}. Einstein suggested that extremely violent events in the universe could
generate ripples in the fabric of space-time that will travel, like electromagnetic waves, at the speed 
of light. However, given the vanishingly small nature of these ripples, Einstein doubted that gravitational 
waves would ever be detected. Fast forward a century and the LIGO-Virgo collaboration has achieved
the unimaginable: the very first direct detection of gravitational waves\,\cite{Abbott:2016}. In this first
instance, the source of the gravitational waves were two colliding black holes, with initial masses of 
$36\,M_{\odot}$ and $29\,M_{\odot}$ and producing nearly 3 solar masses of gravitational-wave 
radiation. Those gravitational waves traveled from a distance of about 400 Mpc to induce a miniscule
gravitational-wave strain amplitude at the two LIGO detectors of $h\!=\!\Delta L/L\!\approx\!10^{-21}$. 
For an interferometer consisting of two $L\!=\!4\,{\rm km}$ arms, this implies changes in the arm length 
of about $\Delta L\!\approx\!4\times\!10^{-3}\,{\rm fm}$, or a tiny fraction of the size of an elementary 
proton. As such, Einstein's skepticism towards the direct detection of gravitational waves hardly comes 
as a surprise. 

This remarkable accomplishment was soon followed by the first direct detection of gravitational 
waves from a binary neutron star inspiral (GW170817)\,\cite{Abbott:2017}. Unlike black holes, the
electromagnetic---and in the future neutrino---emission from these cataclysmic events may also be 
detected. Indeed, nearly two seconds after the gravitational waves were detected by the LIGO-Virgo
interferometers, the Fermi Gamma-ray Space Telescope and the International Gamma-Ray Astrophysics 
Laboratory recorded a short duration $\gamma$-ray burst, confirming the long-held belief that short 
gamma-ray bursts are associated with binary neutron-star mergers. And within eleven hours of these 
detections, both ground- and space-based telescopes identified the associated kilonova, the 
electromagnetic transient believed to be powered by the radioactive decay of the heavy elements 
synthesized in the r-process, thereby answering one of the most fundamental questions in all of science: 
What is the origin of the heavy elements? Thus, in one clean sweep, GW170817 and its associated 
electromagnetic counterparts, launched the new era of multimessenger astronomy.

\begin{figure}[ht]
\smallskip
\centering
\includegraphics[width=0.75\columnwidth]{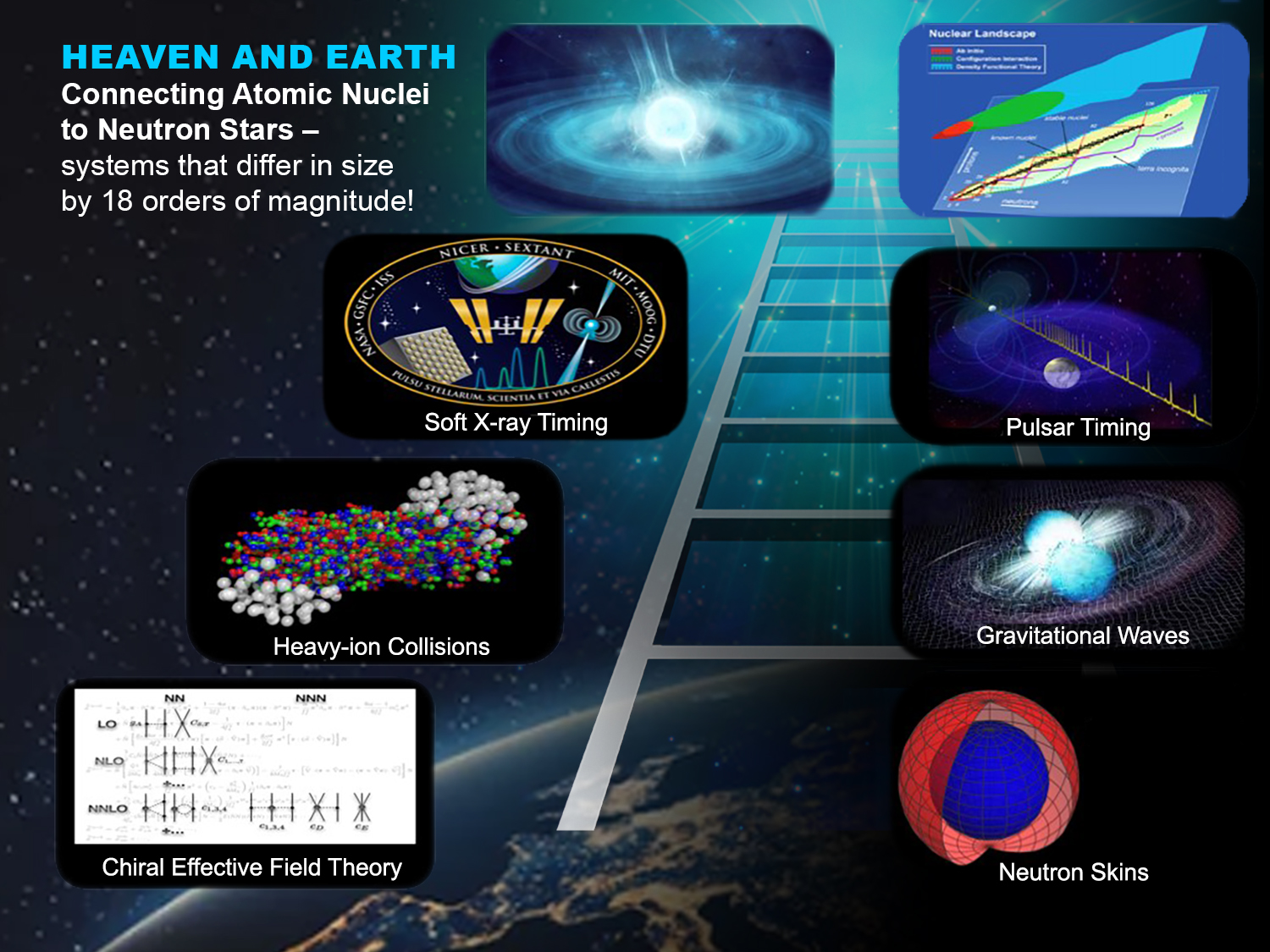}
\caption{The equation of state density ladder. Each rung in the ladder informs the equation of the state of 
              neutron-rich matter in an appropriate density regime. Figure adopted from the 2023 Long Range
              Plan for Nuclear Science\,\cite{LRP:2023}.}
\label{Fig1}       
\end{figure}

Whereas GW170817 has generated a paradigm shift in our understanding of neutron stars, many
 other developments continue to shape our understanding of the structure, dynamics, and composition
 of neutron stars. Notable among them is the determination of the most massive neutron star to 
 date\,\cite{Cromartie:2019kug,Fonseca:2021wxt}; the first simultaneous determination of the 
 mass and radius of neutron stars\,\cite{Riley:2019yda,Miller:2019cac,Riley:2021pdl,Miller:2021qha}; 
 measurements of the thickness of the neutron skin of neutron-rich nuclei\,\cite{Adhikari:2021phr,
 Adhikari:2022kgg}; and significant theoretical advances that incorporate all these discoveries to 
 provide refined new insights and improved predictions with quantified 
 uncertainties\,\cite{Fattoyev:2017jql,Drischler:2020hwi,Reed:2021nqk}. The confluence of so many 
 advances motivates the creation of the so-called equation of state density ladder, akin to the cosmological 
 distance ladder. As illustrated in Fig.\ref{Fig1}\,\cite{LRP:2023}, while the synergy is compelling, no single 
 method can determine the equation of state (EOS) over the entire density range, yet each rung on the 
 ladder informs the EOS in a suitable domain that overlaps with its neighboring rungs. Underscoring the 
 synergy among such seemingly distinct fields is the main goal of this contribution.

\section{GW170817: Tidal imprints on the EOS}
\label{sec:GW170817}

\begin{figure}[ht]
\smallskip
\centering
\includegraphics[width=0.75\columnwidth]{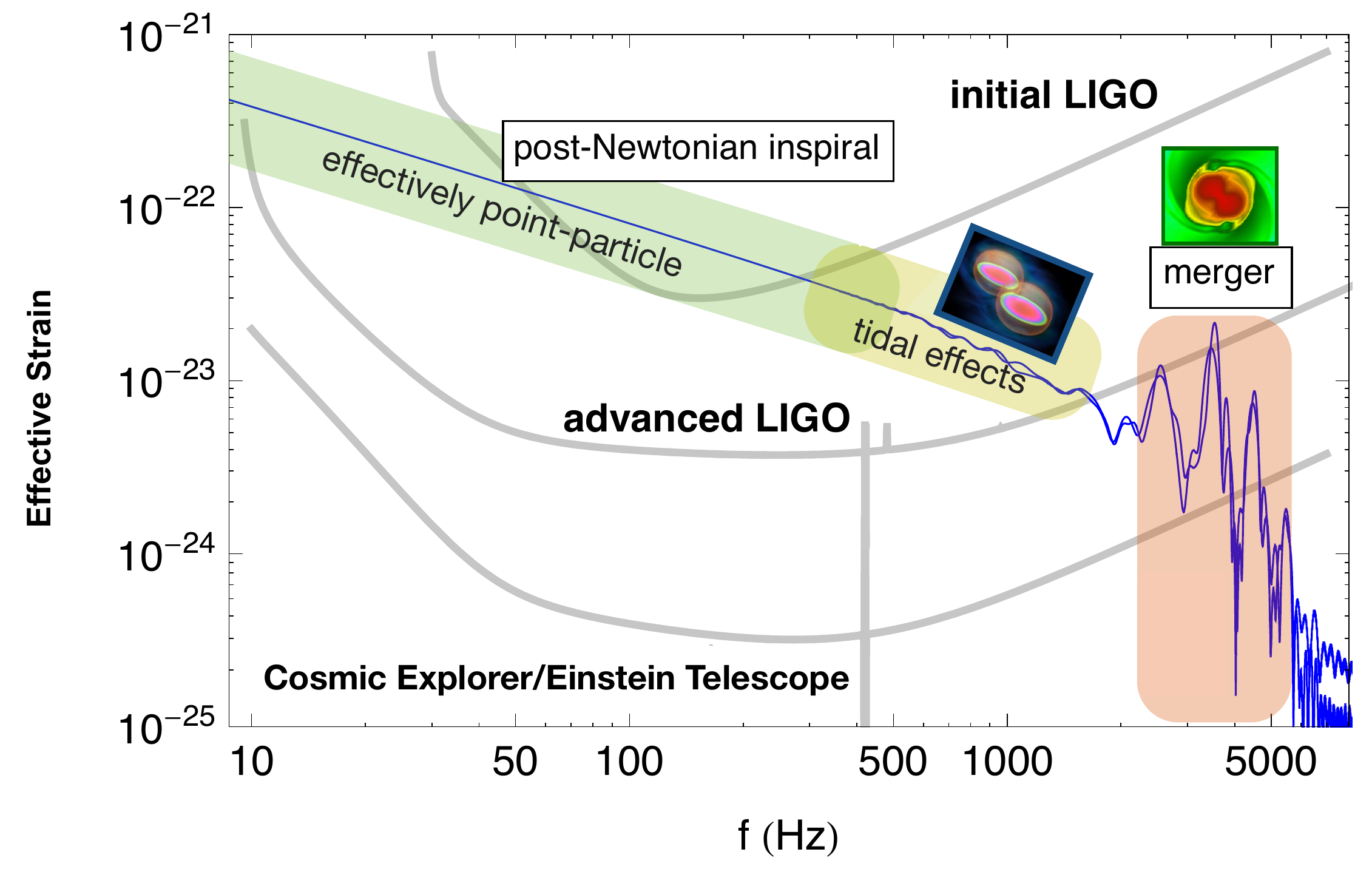}
\caption{Illustration of the gravitational-wave signal emitted during the last stages of the binary neutron star 
	     coalescence, such as in the case of GW170817. At present, the most stringent constraints on the 
	     EOS come from the few-hundred Hz region where tidal effects become important. During the second 
	     observing run, advanced LIGO/Virgo reached the required sensitivity to detect GW170817. Next 
	     generation gravitational wave detectors, such as Cosmic Explorer and the Einstein Telescope, will 
	     reach such a high sensitivity that the detection of binary neutron star mergers will become routine. 
	     Figure adapted from Ref.\,\cite{Andersson:2017}.}
\label{Fig2}       
\end{figure}

We start this section by displaying in Fig.\ref{Fig2} the effective gravitational wave strain $h\!=\!\Delta L/L$ 
as a function of frequency during the last stages of the binary neutron-star inspiral\,\cite{Andersson:2017}.
Below a few $100\,{\rm Hz}$, the gravitational-wave profile is insensitive to matter effects and hence 
indistinguishable from two colliding ``point particles'' (i.e., black holes). Around $400\,{\rm Hz}$, matter effects 
become important as both neutron stars develop a mass quadrupole in response to the tidal fields generated 
by the companion star which, in turn, accelerates the coalescence. At merger, the sensitivity to the EOS remains 
high. However, for such violent event one requires both knowledge of the EOS at finite temperature and the use
of numerical relativity in the strong coupling regime.

The measured gravitational wave strain is related to the intrinsic properties of the binary source as follows:
\begin{equation}
 h(t) = \frac{2G}{c^{4}D}\ddot{Q} \rightarrow 10^{-21}\left(\frac{\rm Mpc}{D}\right),
 \label{h1}
\end{equation}
where $D$ is the distance to the source and $Q$ is the induced, time-varying mass quadrupole. The arrow in the 
above expression indicates the relevant scale in the $f\!\simeq\!400\,{\rm Hz}$ region where the orbital separation 
of the binary system is a few hundred kilometers. In the case of GW170817, with the source at a distance of 
$40\,{\rm Mpc}$, the gravitational wave strain is of the order of $h\!\sim\!10^{-23}$.

Encoded in the gravitational wave signal are matter effects imprinted in the tidal deformability. In the linear 
regime, the induced mass quadrupole $Q_{ij}$ is proportional to the tidal field $\mathcal{E}_{ij}$ created 
by the companion, with the constant of proportionality being the tidal deformability. That 
is\,\cite{Abbott:2017,Hinderer:2009ca},
\begin{equation}
 Q_{ij} \propto \Lambda \mathcal{E}_{ij}, 
 \label{Qij}
\end{equation}
where the dimensionless tidal deformability $\Lambda$ is given by
\begin{equation}
  \Lambda = \frac{2}{3}k_{2}\left(\frac{c^{2}R}{GM}\right)^{5}.
 \label{Lambda}
\end{equation}
Here $k_{2}$ is the dimensionless second Love number\,\cite{Hinderer:2009ca}, and $M$ and $R$ 
are the stellar mass and radius, respectively. The tidal deformability is highly sensitive to the underlying
EOS because it is proportional to the $5^{\rm th}$ power of the stellar compactness. The second Love 
number $k_{2}$ is also sensitive to the EOS, as its value emerges from the solution of a nonlinear, 
first-order differential equation that uses as input the EOS\,\cite{Postnikov:2010yn,Fattoyev:2012uu}. 

Shortly after the publication of the discovery paper\,\cite{Abbott:2017}, a refined analysis of GW170817 by
the LIGO/Virgo collaboration reported a tidal deformability of $\Lambda_{1.4}\!=\!190^{+390}_{-120}$.
Such a relatively small value of $\Lambda_{1.4}$ suggests that neutron stars are fairly compact, thereby 
favoring a soft equation of state\,\cite{Abbott:2018}. Moreover, estimates of the stellar radius of the two 
coalescing stars suggest a common radius of $R_{1}\!=\!R_{2}\!=\!(11.9\pm1.4)\,{\rm km}$, at the 90\%
confidence level.

\section{PSR J0740: Constraints on the EOS at the highest densities}
\label{sec:J0740}

Whereas Fig.\ref{Fig1} underscores the strong interplay among different theoretical, experimental, and 
observational techniques, the ideal---and perhaps only---method to determine the EOS at the highest 
densities found in the neutron-star interior is through the measurement of massive neutron stars. To date, 
PSR J074 is among the most---if not the most---massive neutron star ever recorded\,\cite{Cromartie:2019kug,
Fonseca:2021wxt}. The precise and largely model-independent extraction of the mass of PSR J074 relies 
on measuring the Shapiro delay\,\cite{Shapiro:1964}, a critical test of general relativity. 

\begin{figure}[ht]
\smallskip
\centering
\includegraphics[width=0.8\columnwidth]{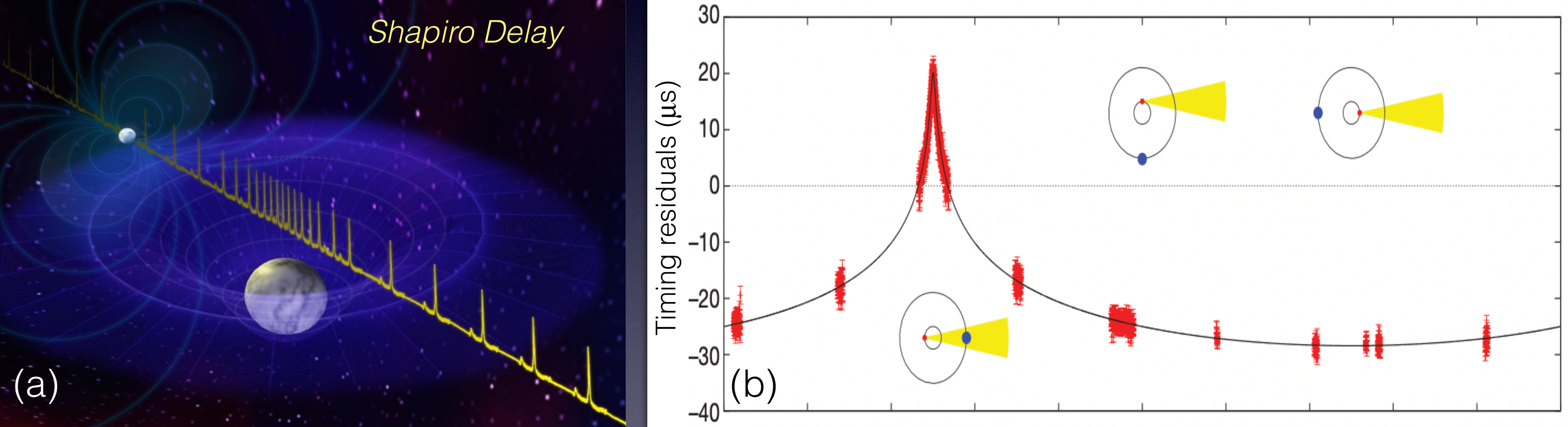}
\caption{(a) Illustration of the Shapiro delay in superior conjunction, when the white-dwarf star is located 
              between the pulsar and the terrestrial observer. (b) Shapiro delay measurement for PSR J1614, as
              displayed by the timing residuals for one entire period. The pulsar is shown in red, the white dwarf
	     companion in blue, and the emitted signal from the pulsar pointing towards the observer in yellow. 
	     Figure adapted from Ref.\,\cite{Demorest:2010bx}.}
\label{Fig3}       
\end{figure}

PSR J074 is part of a binary system composed of the pulsar and a white-dwarf companion. In Newtonian 
gravity, measuring the orbital parameters of the binary system, such as the period and the length of the
semi-major axis, only determines the sum of the individual masses; Shapiro delay helps breaks the mass 
degeneracy. As shown in Fig.\ref{Fig3}, at superior conjunction---when the white dwarf is located between 
the pulsar and the observer---the light wave emitted by the pulsar is delayed in reaching the observer 
because it deeps into the gravitational well generated by the companion star. In the particular case of 
PSR J074, the time delay is given by
\begin{equation}
  \delta t = \frac{2GM_{{}_{\rm WD}}}{c^{3}}\ln\left(\frac{4R_{\star}R_{\oplus}}{d^{\,2}}\right),
 \label{Lambda}
\end{equation}
where $R_{\star}$ is the distance from the white dwarf to the neutron star, $R_{\oplus}$ is the distance 
from the white dwarf to the terrestrial observer, and $d$ is the distance of closest approach of the light 
wave to the white dwarf. Note that if instead the pulsar is located between the white dwarf and the observer, 
then there is no gravitational time delay. As such, once the white-dwarf mass is extracted from the Shapiro 
delay, the mass of the neutron star may be readily inferred from Kepler's third law.

Pulsar timing is a powerful observational technique that accounts for every rotation over long periods of time, 
ensuring high precision in the determination of the individual stellar masses. In the case of PSR J074, the 
Shapiro delay amounted to about $20\,\mu s$\,\cite{Cromartie:2019kug}, resulting in a mass determination
for PSR J074 of $M\!=\!(2.08\pm 0.07)\,M_{\odot}$\,\cite{Fonseca:2021wxt}. 

\section{NICER: Building the Mass-Radius Relation}
\label{sec:NICER}

The Neutron Star Interior Composition Explorer (NICER) was launched in 2017 aboard a SpaceX's Falcon 9 
rocket and deployed to the International Space Station. NICER is dedicated to the study of the structure, dynamics, 
and composition of neutron stars. To constrain the EOS, NICER relies on the powerful technique of Pulse Profile 
Modeling to monitor electromagnetic emission from the hot spots located on the surface of the neutron 
star\,\cite{Psaltis:2013fha,Watts:2016uzu}. As the neutron star spins, the hot spots come in and out of 
view producing periodic variations in the brightness, namely, a pulse profile. Because of gravitational light-bending,
x-rays emitted from the ``back of the star'' are detected by the powerful NICER instruments. Because the
amount of light bending is highly sensitive to the stellar compactness, NICER provides critical information 
on the mass-radius relation, often referred to as the holy grail of neutron star physics. 
\begin{figure}[ht]
\smallskip
\centering
\includegraphics[width=0.5\columnwidth,height=6cm]{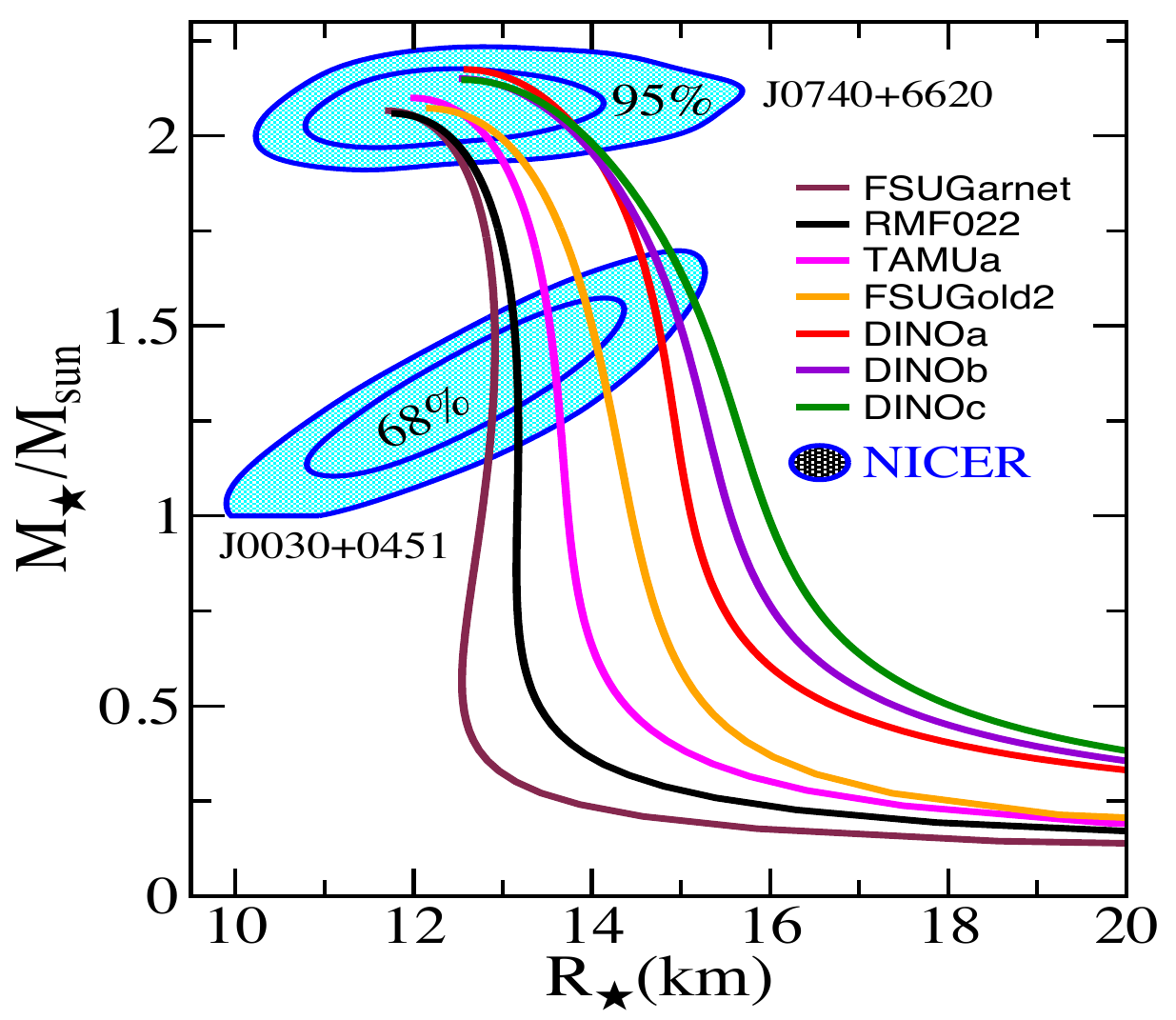}
\caption{The holy grail of neutron star physics: The mass-radius relation. Predictions for a collection of 
covariant energy density functionals\,\cite{Reed:2023cap} are displayed alongside the two sets of 
contours (at 68\% and 95\% confidence) extracted by the NICER collaboration on the mass and radius 
of the two millisecond pulsars PSR J0030\,\cite{Riley:2019yda,Miller:2019cac}  and 
PSR J0740\,\cite{Riley:2021pdl,Miller:2021qha}. Figure adopted from Ref.\,\cite{Reed:2023cap}.}
\label{Fig4}       
\end{figure}

Before the deployment of NICER, simultaneous knowledge of the mass and radius of even a single 
neutron star was not possible. Since then, masses and radii are now available for two millisecond pulsars: 
PSR J0030\,\cite{Riley:2019yda,Miller:2019cac} and PSR J0740\,\cite{Riley:2021pdl,Miller:2021qha}. 
We display in Fig.\ref{Fig4} the mass-radius relation as predicted by a collection of covariant energy density 
functionals\,\cite{Reed:2023cap}. Also shown are contour plots obtained from the NICER analysis of the
two millisecond pulsars. Highly relevant is that PSR J0740 is the same heavy pulsar whose mass was 
already accurately determined using pulsar timing; see Sec.\ref{sec:J0740}. For this massive pulsar, the 
average stellar radius is about 12.4\,km\,\cite{Riley:2021pdl,Miller:2021qha}, suggesting that in the density 
regime sensitive to stellar radii---about two-to-three times nuclear saturation---the EOS is relatively stiff. 

\section{Heaven and Earth: Neutron skins and neutron stars}
\label{sec:NSkins}

The neutron skin thickness of an atomic nucleus is defined as the difference in radii between the neutron 
and proton densities. It is a remarkable fact that despite a difference in length scales of about 18 orders of 
magnitude, the thickness of the neutron skin and the radius of a neutron star are strongly 
correlated\,\cite{Horowitz:2000xj,Horowitz:2001ya}. Whereas the excess neutrons in a neutron-rich nucleus 
push against surface tension, the neutrons in a neutron star push against the pull of gravity. Hence, both 
atomic nuclei and low-mass neutron stars probe similar regions of the equation of state\,\cite{Carriere:2002bx}. 
However, low mass neutron stars have been difficult to  find. Thus, the neutron skin thickness may serve as 
a proxy for the radius of low mass neutron stars, thereby providing an ideal observable to constrain the EOS 
in the vicinity of nuclear saturation density. 

In quantifying the impact of excess neutrons in both atomic nuclei and neutron stars, the nuclear symmetry 
energy plays a pivotal role. At zero temperature, as it is appropriate for isolated neutron stars, the energy per nucleon 
depends only on the individual neutron and proton densities. Alternatively, one may express the binding energy per 
nucleon in terms of the total baryon density $\rho\!=\!\rho_{n}\!+\!\rho_{p}$ and the neutron-proton asymmetry 
$\alpha\!=\!(\rho_{n}\!-\!\rho_{p})/\rho$. That is,
\begin{equation}
  {\mathcal E}(\rho,\alpha)={\mathcal E}_{\rm SNM}(\rho) + \alpha^{2}{\mathcal S}(\rho) + {\mathcal O}(\alpha^{4}) \,,
 \label{EoverA}
\end {equation}
where the leading term represents the energy per nucleon of symmetric nuclear matter and the first-order 
correction to the symmetric limit is the nuclear symmetry energy ${\mathcal S}(\rho)$. No odd terms in $\alpha$ 
appear because in the absence of electroweak interactions it is equally costly to have excess neutrons than excess 
protons. From the above expression one deduces that the nuclear symmetry energy is defined as 
\begin{equation}
  {\mathcal S}(\rho) = \frac{1}{2} \left[\frac{\partial^{2}{\mathcal{E(\rho,\alpha)}}}{\partial\alpha^{2}}\right]_{\alpha=0}
   \!\!\approx \Big({\mathcal E}_{\rm PNM}(\rho) - {\mathcal E}_{\rm SNM}(\rho)\Big),
 \label{SymmE}
\end {equation}
where ${\mathcal E}_{\rm PNM}(\rho)\!=\!{\mathcal E}(\rho,\alpha\!=\!1)$ is the energy of pure neutron matter
and the last expression is obtained by neglecting all ${\mathcal O}(\alpha^{4})$ terms in the expansion.

\begin{figure}[ht]
\smallskip
\centering
\includegraphics[width=0.8\columnwidth]{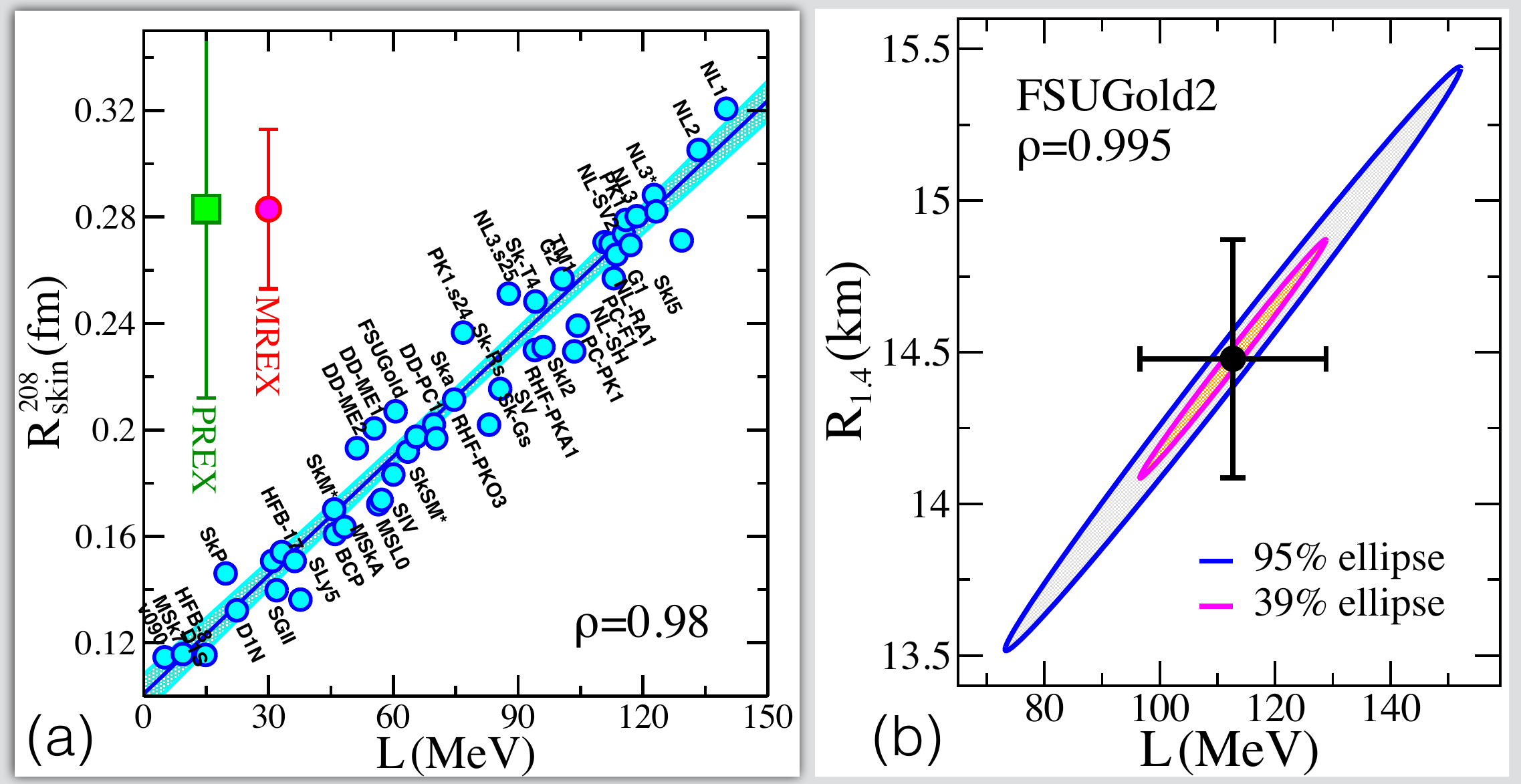}
\caption{(a) Neutron skin thickness of ${}^{208}$Pb against the slope of the symmetry energy $L$ as 
predicted by a large set of energy density functionals\,\cite{RocaMaza:2011pm}. The error bands
provide a best-fit to the model predictions and reveal a very high correlation coefficient of $\rho\!=\!0.98$.
Also shown are results from the PREX campaign\,\cite{Adhikari:2021phr} and from a future
experiment (MREX) that promises a factor of two improvement in precision. The assumed central value 
for MREX coincides with the extracted PREX value of $R_{\rm skin}^{208}\!=\!0.283\,{\rm fm}$. Figure 
adapted from Ref.\cite{RocaMaza:2011pm}. (b) Neutron radius of a $1.4\,M_{\odot}$ neutron star against 
the slope of the symmetry energy $L$ as predicted by the FSUGold2 energy density 
functional\,\cite{Chen:2014sca}. Together with the theoretical error bars, also shown are 39\% and 95\% 
confidence ellipses that suggest a very tight correlation coefficient of nearly one between these two observables. 
Figure adapted from Ref.\cite{Yang:2019fvs}.}
\label{Fig5}       
\end{figure}

The left-hand panel in Fig.\ref{Fig5} shows the strong correlation between the neutron skin thickness of 
${}^{208}$Pb and the slope of the symmetry energy $L$, a quantity that is closely related to the pressure 
of pure neutron matter at saturation density. The figure displays predictions from a large collection of 
non-relativistic and relativistic energy density functionals for both $R_{\rm skin}^{208}$ and 
$L$\,\cite{RocaMaza:2011pm}. The strong correlation is reflected in a large correlation coefficient of 
$\rho\!=\!0.98$. Also shown in the figure are the results from the PREX analysis\,\cite{Adhikari:2021phr} 
and from the future Mainz Radius EXexperiment (MREX) that promises a factor of two improvement in the 
determination of $R_{\rm skin}^{208}$ relative to PREX\,\cite{Mammei:2023kdf}. 

The right-hand panel of Fig.\ref{Fig5} displays another strong correlation between the slope of the symmetry 
energy $L$ and the radius of a $1.4\,M_{\odot}$ neutron star---an object that is typically 18 orders of magnitude 
larger than an atomic nucleus. In this case, a statistical analysis has been carried out using only the FSUGold2 
density functional\,\cite{Chen:2014sca}. With a correlation coefficient of nearly one, $R_{\rm skin}^{208}$ 
provides a powerful proxy for $L$, that ultimately enables the emergence of a powerful data-to-data relation 
between $R_{\rm skin}^{208}$ and the radius of (low-mass) neutron stars\,\cite{Carriere:2002bx,Piekarewicz:2019ahf}.

\section{Conclusions and Outlook}
\label{sec:Conclusions}

The main goal of the present contribution was to highlight the remarkable synergy between seemingly
unrelated fields in the brand new era of multimessenger astronomy. During the last few years, gravitational 
wave detections from binary mergers, novel electromagnetic observations that constrain neutron star 
masses and radii, and terrestrial experiments that serve as proxy for the determination of stellar radii 
of low mass stars, have all been instrumental in transforming our understanding of neutron stars. 

\begin{figure}[ht]
\smallskip
\centering
\includegraphics[width=0.75\columnwidth]{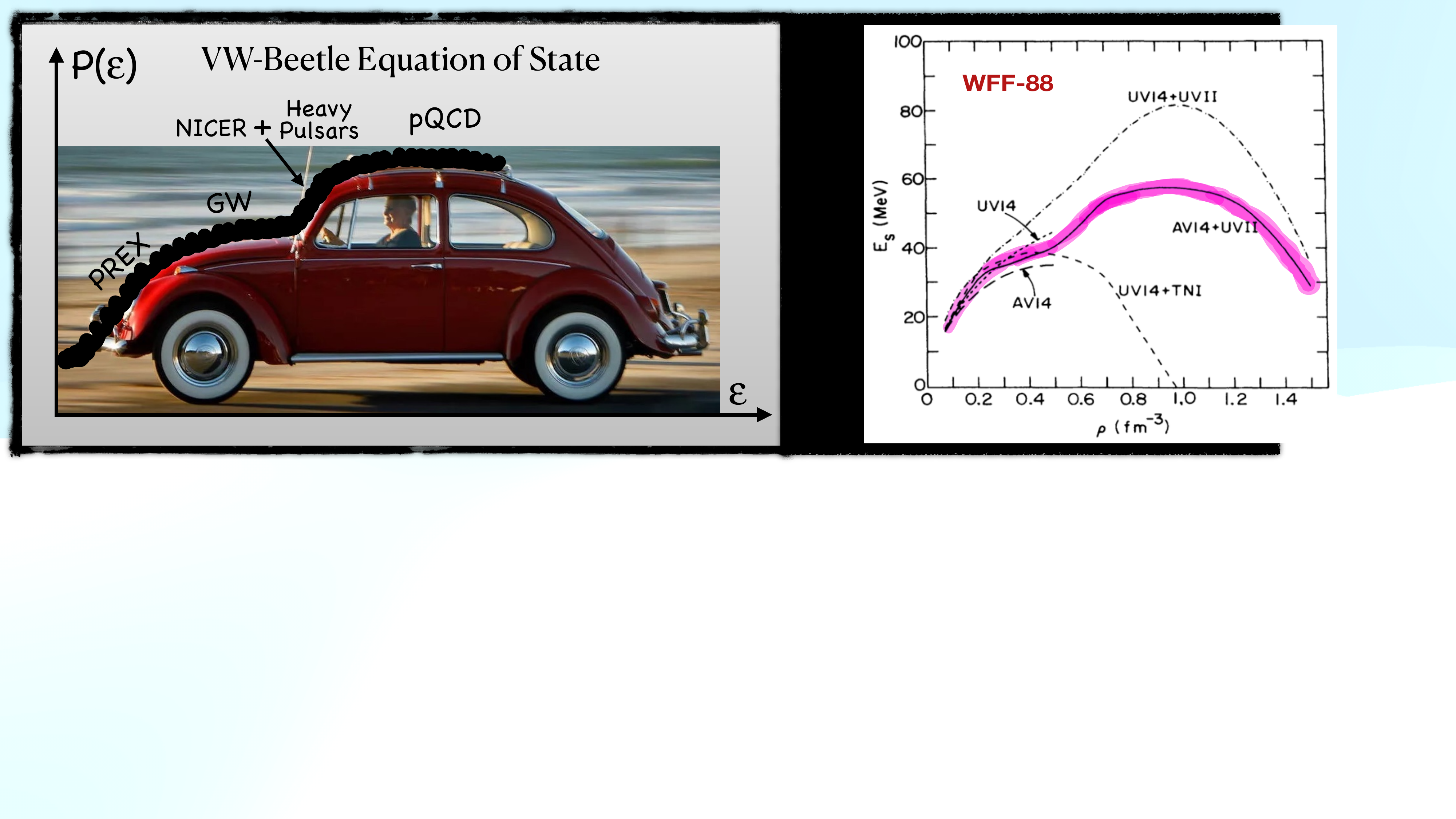}
\caption{Schematic illustration of the equation of state, namely, pressure as a function of the energy 
 density, deduced from incorporating information from terrestrial experiments and astrophysical 
 observations.}
\label{Fig6}       
\end{figure}

Although much of the alluded synergy is captured by the EOS density ladder displayed in Fig.\ref{Fig1}, 
a quantitative picture is staring to emerge. According to PREX, the extraction of a thick neutron skin 
suggests that the EOS in the vicinity of nuclear saturation density is stiff, namely, that the pressure
increases rapidly with increasing density. At the slightly higher densities probed in the interior of a 
$1.4\,M_{\odot}$ neutron star, the small tidal deformability reported by the LIGO-Virgo collaboration 
indicates that neutron stars are compact, implying a softening of the symmetry energy. However, 
stellar radii reported by the NICER mission as well as the observation of heavy neutron stars require 
that the EOS stiffens again at the highest densities probed in the stellar core. Such unique evolution 
from stiff to soft and back to stiff may reflect non-trivial dynamics, perhaps indicative of an exotic phase
transition in the stellar interior. As displayed in Fig.\ref{Fig6}, such unique EOS profile is highly 
reminiscent of the conspicuous Volkswagen beetle that roamed the streets of Mexico City for 
many years. Although perhaps just an interesting curiosity, it is worth mentioning that the nuclear 
symmetry energy predicted back in 1988 by Wiringa, Fiks, and Fabrocini using the AV14+UVII
Hamiltonian displays precisely the same shape; see Fig.8 in Ref.\,\cite{Wiringa:1988}.

If the present is bright the future is even brighter. MREX will measure the neutron skin thickness
of ${}^{208}$Pb with improved precision relative to PREX. As such, MREX will provide stringent 
constrains on the slope of the symmetry energy and ultimately on the EOS of neutron rich matter. In 
the heavens, the European Space Agency Athena X-ray mission will be able to detect fainter sources 
and reduce statistical errors that will provide simultaneous mass-radius measurements for a large 
number of neutron stars\,\cite{Nandra:2013}. In turn, with gravitational-wave astronomy opening
a brand new window into the universe, next-generation gravitational wave observatories, such 
as the Cosmic Explorer and the Einstein Telescope, will be able to detect binary neutron star 
mergers throughout the entire observable universe. Indeed, third-generation gravitational wave 
observatories will detect countless neutron star mergers, enabling the precise determination of stellar 
radii with an unparalleled precision of 100 meters, or about one part in a hundred\,\cite{Evans:2021gyd}. 
Such unprecedented measurements are poised to offer a first glimpse at the composition of neutron 
stars and at the unique possibility of inferring the existence of new exotic states of matter in the stellar 
interiors. 

\section*{Acknowledgments}
 This material is based upon work supported by the U.S. Department of Energy 
 Office of Science, Office of Nuclear Physics under Award Number DE-FG02-92ER40750.




\end{document}